\documentclass[aps,prl,superscriptaddress,floatfix]{revtex4}
\usepackage{amsfonts}
\usepackage{epsfig}
\usepackage{amsmath,amssymb}
\usepackage{times}
\usepackage{xcolor}
\usepackage{multirow}

\begin{document}

\title{Large epidemic thresholds emerge in heterogeneous networks of heterogeneous nodes}

\author{Hui Yang\footnote{Correspondence to: jenifferyang25@gmail.com}} 
\affiliation{Web Sciences Center, University of Electronic
Science and Technology of China, Chengdu 610054, China}
\affiliation{Department of Engineering Mathematics, University of Bristol, Bristol BS8 1UB, UK}

\author{Ming Tang\footnote{Correspondence to: tangminghuang521@hotmail.com}}
\affiliation{Web Sciences Center, University of Electronic
Science and Technology of China, Chengdu 610054, China}

\author{Thilo Gross}
\affiliation{Department of Engineering Mathematics, University of Bristol, Bristol BS8 1UB, UK}

\date{\today}

\begin{abstract}
One of the famous results of network science states that networks with heterogeneous connectivity are more susceptible 
to epidemic spreading than their more homogeneous counterparts. 
In particular, in networks of identical nodes it has been shown that network heterogeneity, i.e.~a broad degree distribution, can lower the epidemic threshold at which 
epidemics can invade the system. Network heterogeneity can thus allow diseases with lower transmission probabilities
to persist and spread.   
However, it has been pointed out that networks in which the properties of nodes are intrinsically heterogeneous can be very resilient to disease spreading. 
Heterogeneity in structure can enhance or diminish the resilience of networks with heterogeneous nodes, depending on the correlations between the topological and intrinsic properties. 
Here, we consider a plausible scenario where people have intrinsic differences in susceptibility and adapt their social 
network structure to the presence of the disease. 
We show that the resilience of networks with heterogeneous connectivity can surpass those of networks with homogeneous connectivity.
For epidemiology, this implies that network heterogeneity should not be studied in isolation, it is instead the heterogeneity of infection risk that determines the likelihood of outbreaks. 
\end{abstract}

\maketitle

In the exploration of complex systems, epidemiology plays an important role as a source for toy models and case studies, but also an area where a real world impact can be made \cite{vespignani2012modelling,barrat2008dynamical, hufnagel2004forecast,brockmann2005dynamics, colizza2007predictability, janssen2006toward}. It has been pointed out that new diseases have emerged whenever environmental change brought humans in contact with new pathogen or disease vectors, i.e.~animal hosts of a given disease \cite{karlen1996man}. The past decades have brought rapid environmental change, a growing world population, and increasing long-range connectivity in relevant networks, due to human travel and livestock transports \cite{brockmann2013hidden}. Together with decreasing vaccination levels and misuse of antibiotics, this has led to both emergence of new diseases and the return of old ones, sometimes in the form of highly resistant strains. In the future antibiotics, vaccinations, and quarantine are bound to remain first line of defence against these threats. However, insights from physics that can improve the efficiency of these measures, even if only by a small measure, have the potential to save many lives and relieve the economic burden created by the disease.

Many current studies seek to determine the so-called epidemic threshold, the critical level of infectivity that a pathogen needs to surpass to spread and cause large outbreaks \cite{castellano2010thresholds,boguna2013nature}. This threshold depends on many factors including the structure of the underlying network of contacts, the heterogeneity in the host population, and behavioral responses to the disease.  
Among these, the effect of network structure is perhaps best understood \cite{pastor2001epidemic,colizza2007invasion,parshani2010epidemic, belik2011natural}.
It can be shown that the ability of a disease to spread is generally related to the leading eigenvalue of the networks adjacency or non-backtracking matrices \cite{wang2003epidemic,castellano2010thresholds,hamilton2014tight,karrer2014percolation,rogers2015assessing}. Factors that increase this eigenvalue, lead to lower epidemic thresholds. Two well-known factors that facilitate the spreading of diseases are high network connectivity and network heterogeneity. Here, network heterogenity can be defined as the second moment of the degree distribution, the probability distribution of the number of links on a randomly chosen node. 
In extreme cases, for example in scale-free networks, the epidemic threshold may vanish in the thermodynamic limit such that diseases with arbitrarily low infectivity can still spread \cite{pastor2001epidemic,demirel2012absence}.      

With respect to individual heterogeneity in the host population, some questions remain open. Generally intra-individual and link heterogeneity \cite{yang2012epidemic, wang2014epidemic}, such as different levels of resistance, times of contact, differences in infectitivity or recovery rates, reduce the size and risk of epidemics \cite{miller2007epidemic,miller2008bounding,neri2011heterogeneity,neri2011effect,katriel2012size,smilkov2014beyond}. For instance, it was shown analytically and numerically  that epidemics are most likely if infectivity is homogeneous and least likely if the variance of infectivity is maximized \cite{miller2007epidemic,miller2008bounding}. Comparable results were found in lattice models and in a biological experiment \cite{neri2011heterogeneity}. 
However, heterogeneous susceptibility can make networks more vulnerable to the spread of diseases if the correlation between a node's degree and susceptibility are positive \cite{smilkov2014beyond}. This is intuitive as the intra-individual heterogeniety, in this case, amplifies the negative effect of network heterogeniety. Finally, it was reported that heterogeneous susceptibility can in some scenarios cause a secondary epidemic after a primary outbreak  \cite{katriel2012size,rodrigues2009heterogeneity}.  

The studies referenced above focussed on epidemics on static networks. Another active line or research concerns the effect of dynamic, adaptive \cite{volz2009epidemic,taylor2012epidemic,gorochowski2012evolving,li2010epidemic} or temporal networks \cite{holme2012temporal, cui2014efficient}. While these types of networks are closely related, dynamical networks emphasize the overall statistical effect of changing connectivity \cite{volz2009epidemic,taylor2012epidemic}, adaptive networks emphasize the dynamical response of network structure to the disease state\cite{gross2008adaptive,gross2009adaptive}, and temporal networks focus on the impact of the specific timings of events \cite{perra2012activity, lentz2013unfolding}.

An area that is so far poorly understood is how the behavioral response of individuals to epidemics reshapes the network. Beside institutionalised responses, such as mandatory vaccinations and quarantine, humans react to the outbreak of a major epidemic in a variety of ways, for instance by increasing hygiene, using protective  measures such as face masks, reducing contact with infected individuals \cite{zhang2014suppression}, or avoiding contacts with other humans when infected. The common element in these responses is a reduction in the frequency of contacts between infectious and susceptible individuals. This limits the disease propagation by decreasing the effective network connectivity. However, the ultimate effect of behavioral responses cannot be understood as a static reduction of connectivity. If individuals respond to the epidemic state of other individuals by altering their interactions then a complex dynamical feedback loop is formed in which the state of nodes affects the evolution of the topology, while the topology governs the dynamics of the nodes' state. Thus an \emph{adaptive network} is created.          

In the context of epidemics it has been shown in a simple model \cite{gross2006epidemic} that trying to avoid contact with infected is highly effective against small outbreaks but less effective against an established epidemic. Instead of a single epidemic threshold, such models possess an invasion threshold for new diseases and a (lower) persistence threshold for established epidemics. Thus, a parameter region is formed where a disease cannot spread when it is newly introduced, but can persist when it is already established. Physically speaking, a hysteresis loop is formed and the percolation transition at the onset of the epidemic becomes discontinuous. Under certain conditions the behavioural feedback loop can also lead to the emergence of epidemic cycles \cite{gross2006epidemic,gross2008adaptive}. 

The analysis of wide variety of related models showed that these observations are robust over a wide class of models \cite{zanette2008infection,risau2009contact,janssen2006toward,shaw2008fluctuating,schwartz2010rewiring,wang2011epidemic}. Furthermore, studies showed that behavioral response can increase the impact of targeted vaccinations \cite{shaw2010enhanced}, and that the timing of interventions can be more important than in static networks \cite{yang2012efficient}.      
 
In this paper, we study the combined effect of heterogenity in intra-individual parameters and the behavioral response.
We show that, starting from a well-mixed network, a heterogeneous connectivity is formed. 
It is known that heterogeneous networks of heterogeneous nodes can be very resistant to disease outbreaks,
if certain correlations are present\cite{smilkov2014beyond}. Here we show that these correlations naturally arise in the adaptive network, and that the 
resulting network configuration is generally significantly more resistant to outbreaks than a network with homogeneous topology. Our analysis suggests that the decisive property governing disease invasion is not network heterogeneity but the heterogeneity of the effective disease risk of agents. 

\section*{Results}

\emph{\textbf{Heterogeneous adaptive SIS model}}.
We consider a network of $N$ agents connected by $K$ bidirectional links. We distinguish two types of agents, which we denote as type A and type B, which differ by their resistance to the disease. The type is an internal property of the agent that does not change. Hence, the proportion of agents of type A, $p_{\rm a}$, and the proportion of agents of type B, $p_{\rm b}=1-p_{\rm a}$, are constant in time. For conciseness we denote the total number of agents in a given state $i\in\{{\rm a},{\rm b}\}$
by $N_i=Np_i$. Moreover, nodes carry an additional internal variable that indicates their epidemic state.
We consider a variant of the susceptible-infected-susceptible (SIS) model of epidemic diseases \cite{anderson1991infectious},
such that a given node is either susceptible to the disease, state S, or infected (and infectious), state I. 
The proportion of nodes in the S and I state is denoted as $[S]$ and $[I]=1-[S]$, respectively.

The network is initialized as an Erd\H{o}s-Reny\'{i} random graph, $G(N,K)$. These networks have a narrow, Poissonian degree distribution, such that the network connectivity is homogeneous in the initial state. Each node is randomly assigned a type and 
epidemic state such that desired values of $p_{i}$ and the desired initial values of $[S]$ and $[I]$ are realized. 
Time evolution of the network is then driven by three processes, namely the a) recovery of infected nodes, b) contact avoidance, and c) contagion. These are implemented as follows: a) Infected nodes recover at rate $\mu$, returning to the susceptible state. b) A given link, connecting a susceptible agent to an infected agent is rewired at rate $\omega$. In an rewiring event the original link is cut and a new link between the susceptible node and a randomly chosen other susceptible node is created. c) For every link connecting a susceptible to an infected node, the disease is transmitted along the link at a rate $\beta\psi_i$ that is dependent on the type $i$ of the susceptible node. 

In the following we assume that $\psi_{\rm a}>\psi_{\rm b}$ such that nodes of type A are more susceptible to the disease than nodes of type B. Our mathematical results hold for parameters in arbitrary units, but the rates can be thought of having the dimension of nodes-per-time (recovery) and links-per-time (contagion, rewiring) respectively.   
Throughout the paper we balance the parameters $\psi_{\rm a}$ and $\psi_{\rm b}$ such that $p_{\rm a}\psi_{\rm a} + p_{\rm b} \psi_{\rm b} = \langle \psi \rangle$. The variance of susceptibility is $ \sigma^{2}_{\psi}=(\langle \psi \rangle -\psi_{\rm b}) (\psi_{\rm a}- \langle \psi \rangle)$. 

Most of our results below are found by analytical calculation or continuation of solution branches and are thus non-simulative in nature. However, we compare these results to large agent-based simulations. In these simulations we use an event-driven (Gillespie) algorithm to simulate the stochastic process described above. In simulation we use an Erd\H{o}s-R\'{e}nyi network with $ N=10^{5}$ nodes and $ K=10^{6}$ links with recovery rate $\mu=0.002$, rewiring rate $\omega=0.2$ and average susceptibility $\langle \psi \rangle=0.5$, unless noted otherwise. We vary $\psi_{\rm a}$ as a proxy for heterogeneity. For every choice of $\psi_{\rm a}$ and $\psi_{\rm b}$, the parameter $p_{\rm a}$ is set such that $\langle \psi \rangle=0.5$. All parameters used in simulation runs are stated in the caption of the respective figure. 

\begin{figure}
\begin{center}
\epsfig{file=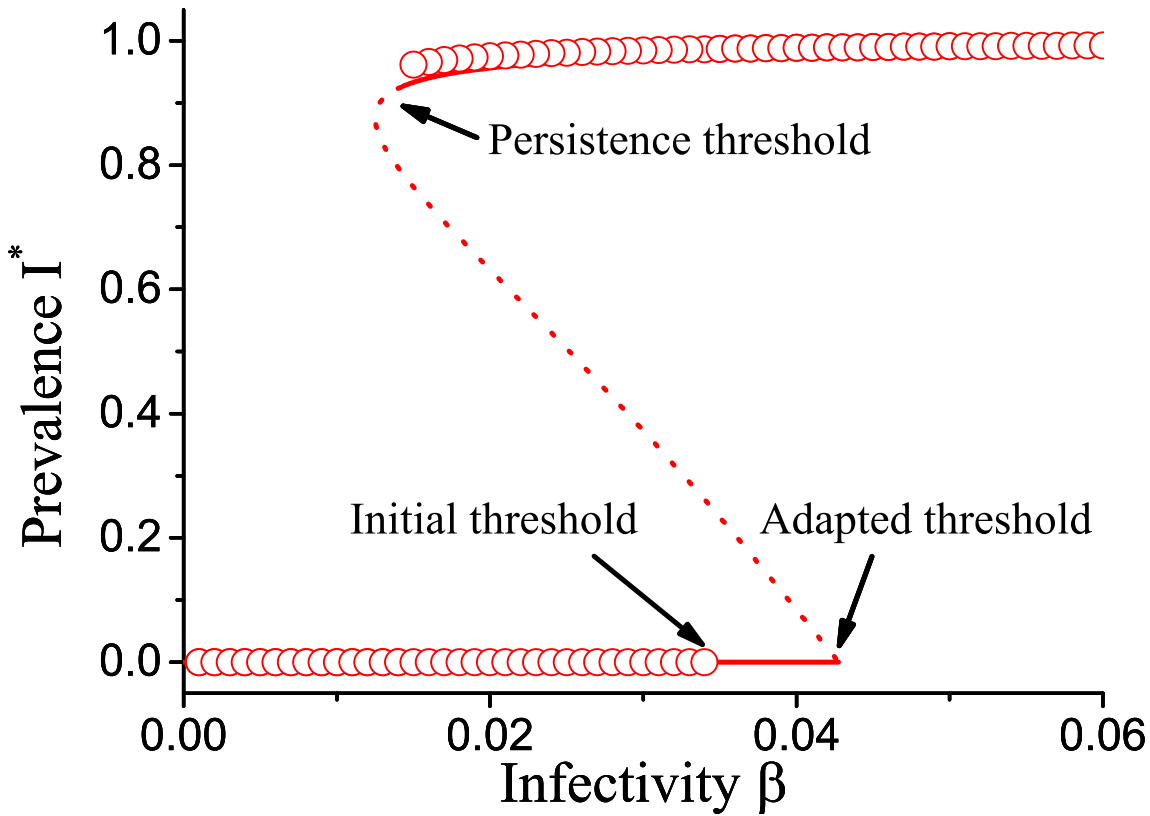,width=1\linewidth}
\caption{(Color online) {\bf Bifurcation diagram of the adaptive heterogeneous SIS model.} Show is the stationary level of disease prevalence $I^*$ as a function of infectivity $\beta$. When the infectivitiy is decreased the endemic state vanishes in a saddle node bifurcation (`Persistence threshold'). The disease free state can be invaded by the epidemic, if the infectivity surpasses a point where a transcritical bifurcation occurs (`Invasion threshold'). Agent-based simulation (circles) and equation-based continuation (lines) provide consistent results on the persistence threshold, but predict different invasion thresholds. This discrepancy appears due to a projection effect, because the state where the disease is extinct is not uniquely defined (see text). In addition to stable solution branches (solid) the continuation also reveals an unstable solution branch (dotted). 
Parameters:  $ \psi_{\rm a} = 0.65 $, $ \psi_{\rm b} = 0.05 $, $ p_{\rm a}= 0.75 $, $\omega=0.2$, $ \mu=0.002 $, $ N=10^{5} $, $ K=10^{6} $.}\label{Fig1}
\end{center}
\end{figure}

\emph{\textbf{Hysteresis in the heterogeneous model}}. To gain some basic intuition, let us first investigate the system by explicit agent-based simulation of the network model \cite{zschaler2013largenet2}. 
For this purpose we evolve the system in time, according to the stochastic rules, until a stationary level of epidemic prevalence is approached. 
Repeating this procedure for different values of infectivity $\beta$ reveals the diagram shown in Fig.~\ref{Fig1}.
The Figure is qualitatively similar to results from the homogeneous adaptive SIS model:
Epidemics starting from a small proportion of initially infected agents go extinct deterministically if the infectivity is below a certain threshold, $\beta_{\rm inv}$, which we identify as the invasion threshold. By contrast, established epidemics, i.e.~simulation runs starting with a higher initial proportion of infected, can persist if the infectivity surpasses a different persistence threshold $\beta_{\rm per}$. 

The persistence threshold is lower than the invasion threshold, such that a bistable region is created.
In this region an epidemic that enters the system at low density goes extinct, but an established large epidemic, that
perhaps entered the system earlier when parameters were different, can persist.  The bistable region constitutes a hysteresis loop: If we slowly increase the infectivity the extinct state is stable until $\beta_{\rm inv}$, where a jump to the endemic state occurs. If we lower $\beta$ again, the system persists in the endemic state up to $\beta_{\rm per}$, where it collapses back down to the extinct state. 

In the following we explore the effect of heterogeneity on the thresholds $\beta_{\rm inv}$ and $\beta_{\rm per}$. To gain analytical insights, we consider a moment expansion of the system \cite{demirel2014moment}. We use symbols of the form $[X_u]$ and $[X_uY_v]$ with $X,Y\in\{{\rm I},{\rm S}\}$ and $u,v\in\{{\rm a},{\rm b}\}$ to respectively denote the proportion of agents and per capita density of links between agents of a given type. For instance $[I_{\rm a}]$ is the proportion of agents that are infected and of type A, and $[S_{\rm a}I_{\rm b}]$ is the per capita density of links between susceptible agents of type A and infected agents of type B. All of these variables are normalized with respect to the total number of nodes $N$. Given the number of infected nodes of a given type we can thus find the number of susceptible nodes by using the conservation law $[I_{u}]+[S_{u}]=p_{u}$. 

The time evolution of the proportion of nodes that are infected and of type A and B  can be respectively written as 
\begin{equation}
\frac{\rm d}{\rm dt}[I_{\rm a}]=-\mu [I_{\rm a}]+\beta \psi_{\rm a} \sum_v [S_{\rm a}I_v],
\end{equation}  
\begin{equation}
\frac{\rm d}{\rm dt}[I_{\rm b}]=-\mu [I_{\rm b}]+\beta \psi_{{\rm b}} \sum_v [S_{\rm b}I_v].
\end{equation}  

For the link densities, using a pair-approximation leads to equations of the form
\begin{equation}
\label{longEq}
\begin{split}
\frac{d[S_{\rm a}S_{\rm a}]}{dt}= & \mu [S_{\rm a}I_{\rm a}] -2\beta\psi_{\rm a} (\frac{[S_{\rm a}S_{\rm a}][S_{\rm a}I_{\rm a}]}{[S_{\rm a}]}+\frac{[S_{\rm a}S_{\rm a}][S_{\rm a}I_{\rm b}]}{[S_{\rm a}]}) 
+\frac{\omega [S_{\rm a}]}{[S_{\rm a}]+[S_{\rm b}]} ([S_{\rm a}I_{\rm a}]+[S_{\rm a}I_{\rm b}]),
\end{split}
\end{equation}
where the terms on the right hand side describe the impact of the different processes on the motif considered, $[S_{\rm a}S_{\rm a}]$ in this example. For instance the first term corresponds to the creation of $S_{\rm a}$-$S_{\rm a}$-links 
due to recovery of the infected node in $S_{\rm a}$-$I_{\rm a}$-links. In total the $I_{\rm a}$-nodes recover at the rate $\mu [I_{\rm a}]$. Every such recovery event creates an expected number of $S_{\rm a}$-$S_{\rm a}$-links that is identical to the average number of $I_{\rm a}$-$S_{\rm a}$-links anchored on an $I_{\rm a}$-node, which is $[I_{\rm a}S_{\rm a}]/[I_{\rm a}]$. In summary, the change in the density of $S_{\rm a}$-$S_{\rm a}$-links due to recovery of $I_{\rm a}$-nodes is $\mu[I_{\rm a}][I_{\rm a}S_{\rm a}]/[I_{\rm a}]=\mu[I_{\rm a}S_{\rm a}]$, which explains the first term in Eq.~(\ref{longEq}). 

In total the moment expansion yields a system of 11 ordinary differential equations. For conciseness we show the remaining equations in the Methods. 

We solve the moment equations by numerical continuation of solution branches using AUTO \cite{doedel2010auto}. 
This reveals branches of stable and unstable steady states (Fig.~1). As in homogeneous adaptive SIS 
model the limits of the hysteresis loop are marked by a fold bifurcation and a transcritical bifurcation point. 
In the fold bifurcation the endemic steady state collides with an unstable saddle and the two states annihilate. 
In the transcritical bifurcation the saddle state intersects the healthy steady state, which causes the healthy state 
to become unstable. The value of $\beta$ in the fold bifurcation point thus marks the persistence threshold $\beta_{\rm per}$ and the critical value in the transcritical bifurcation point marks the invasion threshold $\beta_{\rm inv}$. 

The comparison between the continuation results and agent-based simulation, in Fig.~\ref{Fig1}, shows that both methods are in good agreement regarding the location of the solution branches. Also the values for the persistence threshold agree. However, the continuation predicts a much higher value of the invasion threshold than the agent-based simulation. 

To understand the discrepancy in the thresholds let us again consider the plot in more detail. 
The diagram shows a projection of the full 11-dimensional space spanned by the moment equations. 
In general bifurcation diagrams of this type identify the steady states uniquely. However, this is not
the case when the epidemic is extinct. If there are no infected left, then the dynamics freezes independently 
of the connectivity of the nodes. Thus, as long as $[I_{\rm a}]=[I_{\rm b}]=0$ the state under consideration 
is stationary regardless of the values of $[S_{\rm a}S_{\rm a}]$ and $[S_{\rm b}S_{\rm b}]$. Hence the zero line in the 
bifurcation diagram is really a 2-dimensional plane of absorbing steady states. 

We can now resolve the discrepancy between the continuation and the simulation results. Continuation of the unstable 
solution branch leads to a specific point on the plane of extinct states. This landing point is uniquely defined 
and marks the invasion threshold of the network with the corresponding values of  $[S_{\rm a}S_{\rm a}]$ and $[S_{\rm b}S_{\rm b}]$. However, these values are not identical to those considered in the numerical simulation.     

We argue that both of the invasion thresholds have significance. The simulation is valid for the well-mixed initial 
system, where the network structure has not yet adapted to the presence of the disease. Thus this threshold is relevant 
in case of the arrival of a new disease. By contrast the threshold found by continuation corresponds to a case where the 
network structure has adapted to the disease, for instance due to repeated exposure to same or similar pathogens. In the following, we therefore refer to the two thresholds as the initial and adapted invasion threshold, respectively. 

\begin{figure}
\begin{center}
\epsfig{file=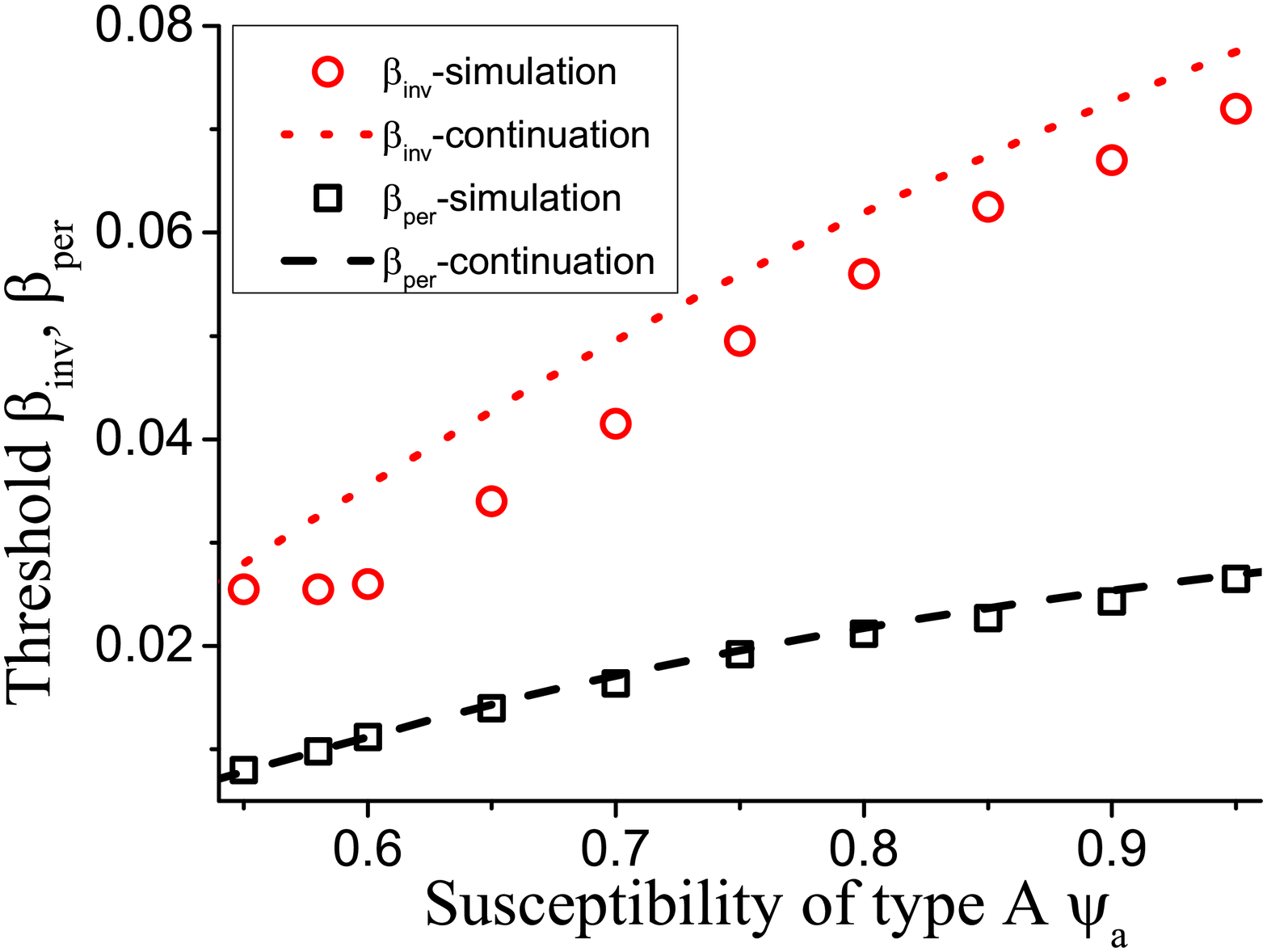,width=1\linewidth}
\caption{(Color online) {\bf Comparison of thresholds.} The plot shows a very good agreement agreement between equation-based continuation (lines) and agent-based simulations (symbols) for the persistence thresholds $\beta_{\rm per}$ (box, dashed). However, a notable difference exists for the invasion thresholds $\beta_{\rm inv}$ (circle, dotted). Parameters: $ \psi_{\rm b} = 0.05 $, $\omega=0.2$, $ \mu=0.002 $, $ N=10^{5} $, $ K=10^{6} $.}\label{Fig2}
\end{center}
\end{figure}

\emph{\textbf{Invasion thresholds and heterogeneity}}.
We emphasize that the observed discrepeancy between the initial and the adapted invasion threshold could not appear in networks of identical nodes. For identical nodes the extinct state is unique on the level of the pair approximation 
($[I]=[II]=[SI]=0$), and thus both thresholds must coincide. The results in Fig.~\ref{Fig2} show that is indeed the case, 
while different thresholds are observed in all networks with heterogeneous nodes.  

\begin{figure}
\begin{center}
\epsfig{file=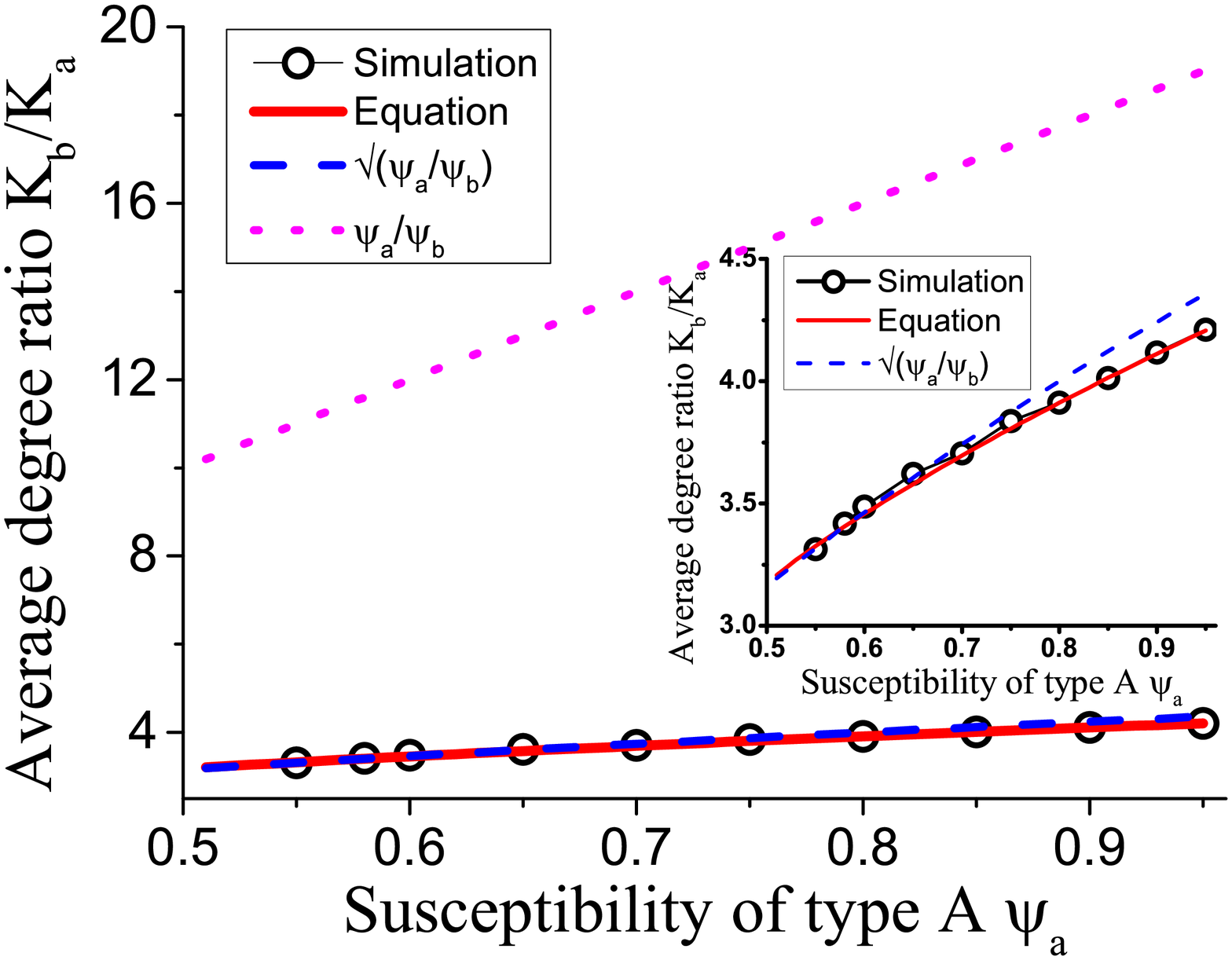,width=1\linewidth}
\caption{(Color online) {\bf Network heterogeneity in the adapted state, indicated by the degree ratio $k_{\rm b}/k_{\rm a}$.}
The dependence of the degree ratio in agent-based simulations (black circles) closely follows the prediction from integration of the ODE model (solid red line), a very good approximation is also provided by the relationship $k_{\rm b}/k_{\rm a} = \sqrt{\psi_{\rm a}/\psi_{\rm b}}$ (blue dashed line), whereas the naive expectation $k_{\rm b}/k_{\rm a} = \psi_{\rm a}/\psi_{\rm b}$ (pink dotted line) overestimates the network heterogeneity significantly. Contrary to expectations 
the networks following the naive solution (the most heterogeneous case), would be maximally stable against disease invasion. Parameters: $ \psi_{\rm b} = 0.05 $, $\omega=0.2$, $ \mu=0.002 $, $ N=10^{5} $, $ K=10^{6} $. Inset: the magnification of the superimposed part of the main figure.}\label{Fig3}
\end{center}
\end{figure}  

We note that the adapted invasion threshold is always higher than the initial threshold. Thus adaptation increases the robustness of the system to disease invasion. Let us therefore explore the adaptation in more detail. The adaptation 
is driven by the rewiring which means that nodes that are frequently infected in average lose links, while 
nodes that are rarely infected gain links. On the population level this means that the average degree of 
nodes of type A, $k_{\rm a}$, and the average degree of nodes of type B, $k_{\rm b}$ change dynamically in response to the 
average proportion of nodes of each type that are infected. Before trying to compute the ratio $k_{\rm b}/k_{\rm a}$ let us first point out that a lower bound is $k_{\rm b}/k_{\rm a}=1$, in this case the degrees are equal, and thus nodes of type A would be infected more frequently, due to their higher susceptibility. Hence $k_{\rm a}$ would decrease and the ratio $k_{\rm b}/k_{\rm a}$ would increase. An upper bound is provided by $k_{\rm b}/k_{\rm a}=\psi_{\rm a}/\psi_{\rm b}$. In this case, $\psi_{\rm a}k_{\rm a}=\psi_{\rm b}k_{\rm b}$ implies the that both types of nodes get infected at equal rate. Because the degree of nodes of type B is higher than the degree of nodes of type A, an infected node of type B will lose links more rapidly and hence the ratio $k_{\rm b}/k_{\rm a}$ will decrease. 

The numerical value of the degree ration $k_{\rm b}/k_{\rm a}$ is shown in Fig.~\ref{Fig3}. To gain also an analytical understanding we resort to a description of the system that is coarser-grained than the full moment expansion. 
First, note that, on a population level $k_i$, the mean degree of nodes of type $i\in\{{\rm a},{\rm b}\}$, obeys a differential equation of the form
\begin{equation}
\frac{\rm d}{{\rm d}t} k_{i} = - k_{i} [I_{i}] u + [S_{i}] v,
\end{equation}
where the two terms denote rewiring losses and gains, and auxialliary variables $u$ and $v$ have been defined to contain all other factors which do not depend on the index $i$. In the steady state we find 
\begin{equation}
k_{i} [I_{i}] = [S_{i}] \frac{v}{u}.
\end{equation}
Thus, when we compute the ratio 
\begin{equation}
\label{almostthere}
\frac{k_{\rm b} [I_{\rm b}]}{k_{\rm a} [I_{\rm a}]} = \frac{[S_{\rm b}]}{[S_{\rm a}]}
\end{equation}
the factors $u$ and $v$ vanish. 
Using a mean field approximation, the epidemic state variables follow equations of the form 
\begin{equation}
\frac{\rm d}{{\rm d}t} [S_{i}] = - q k_{i} \psi_i [S_i] + r [I_i] 
\end{equation} 
where the terms capture the effects of contagion and recovery, respectively, and again auxiliary variables $q$ and $r$ have been defined that contain all other factors that don't explicitly depend on $i$. We use the same trick as before and consider the steady state, where    
\begin{equation}
q k_{i} \psi_i [S_i] = r [I_i] 
\end{equation} 
and hence 
 \begin{equation}
\frac{k_{\rm b} \psi_{\rm b} [S_{\rm b}]}{k_{\rm a} \psi_{\rm a} [S_{\rm a}]} =  \frac{[I_{\rm b}]}{[I_{\rm a}]}. 
\end{equation}    
Substituting this result into Eq.~(\ref{almostthere}) we find 
\begin{equation}
\frac{{k_{\rm b}}^2 \psi_{\rm b}}{{k_{\rm a}}^2  \psi_{\rm a}} \frac{[S_{\rm b}]}{[S_{\rm a}]} = \frac{[S_{\rm b}]}{[S_{\rm a}]}
\end{equation}
and hence 
\begin{equation} 
\frac{k_{\rm b}}{k_{\rm a}}=\sqrt{\frac{\psi_a}{\psi_b}}.
\end{equation}
The result of this mean field argument is in good agreement with numerical results (Fig.~3).
It implies that the rewiring mechanism considered drives the system to a state where the less susceptible nodes (type B) have a higher mean degree, which partly, but not fully, compensates for their lower susceptibility. Thus in the adapted network the nodes of type B get infected more often than they would in a network with homogeneous degree distribution, but still less often than the nodes of type A.  

\begin{figure}
\begin{center}
\epsfig{file=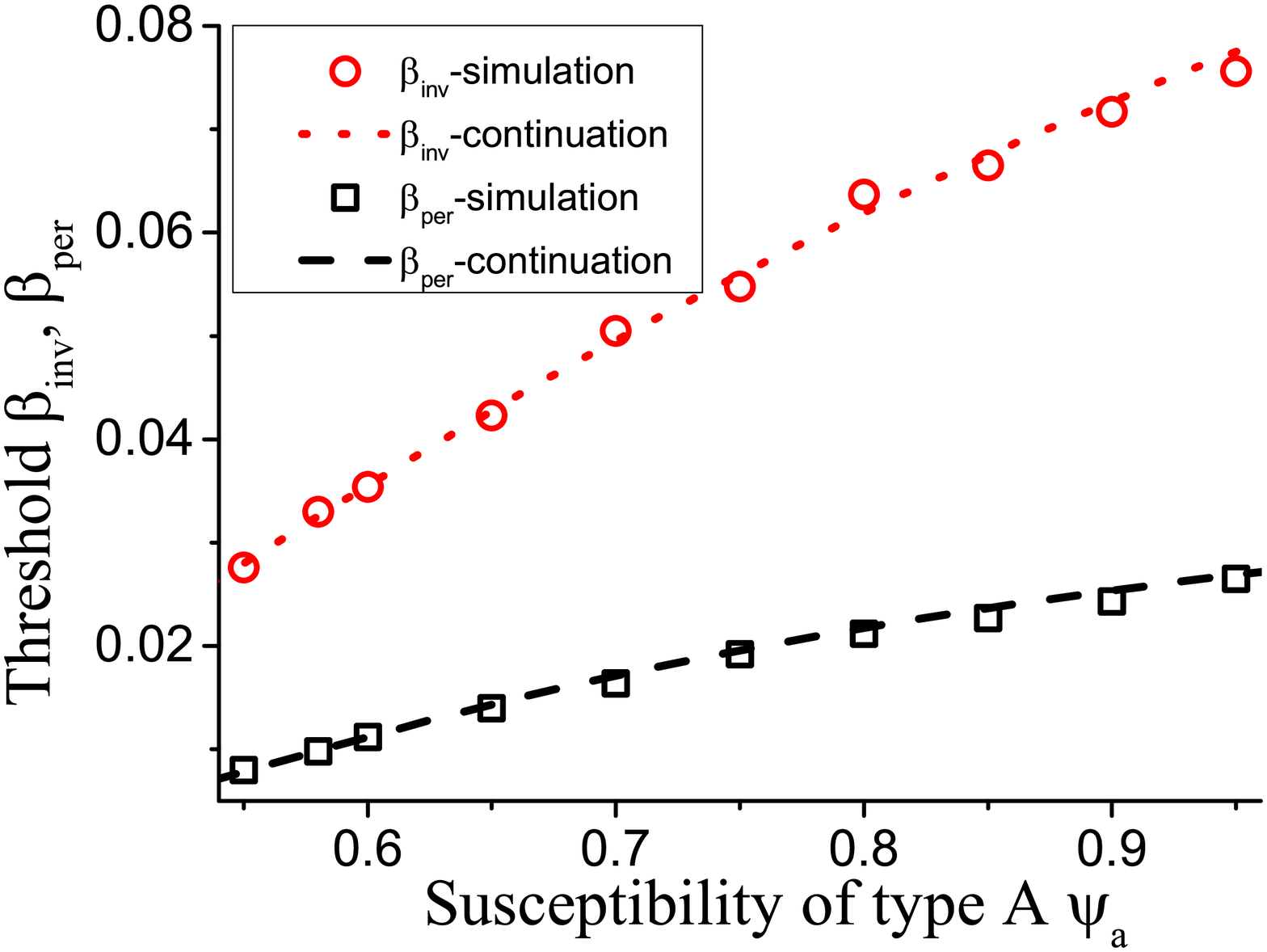,width=1\linewidth}
\caption{(Color online) 
{\bf Comparison of thresholds in self-organized networks.} The plot shows a very good agreement between equation-based continuation (lines) and agent-based simulations started in an artificially created adapted state (symbols) for both the invasion thresholds $\beta_{\rm inv}$ (circle, dotted) and the persistence thresholds $\beta_{\rm per}$ (box, dashed). See Fig.~2 for comparison. Parameters: $ \psi_{\rm b} = 0.05 $, $\omega=0.2$, $ \mu=0.002 $, $ N=10^{5} $, $ K=10^{6} $.
}\label{Fig4}
\end{center}
\end{figure}

To verify that the self-organization of the link distribution explains the observed discrepancy between the initial and the adapted invasion threshold, we turn to the agent-based simulation again. However, in this case we start the simulation from an 
artificially created adapted state. To initialize this state we simulate the system with the same set of initial parameters until the system reaches the stationary state. Then we retain the self-organized link pattern, but reassign all epidemic states, such that all agents are susceptible except for 20 initially infected. Then we simulate the system again until it either reaches the endemic state or the epidemic goes extinct. We locate the epidemic threshold by running a series of such simulations and find the point where the probability to reach the disease-free state becomes zero. The epidemic threshold that is thus found coincides with the result from continuation of the equation-based model (Fig.~\ref{Fig4}).

\begin{figure}
\begin{center}
\epsfig{file=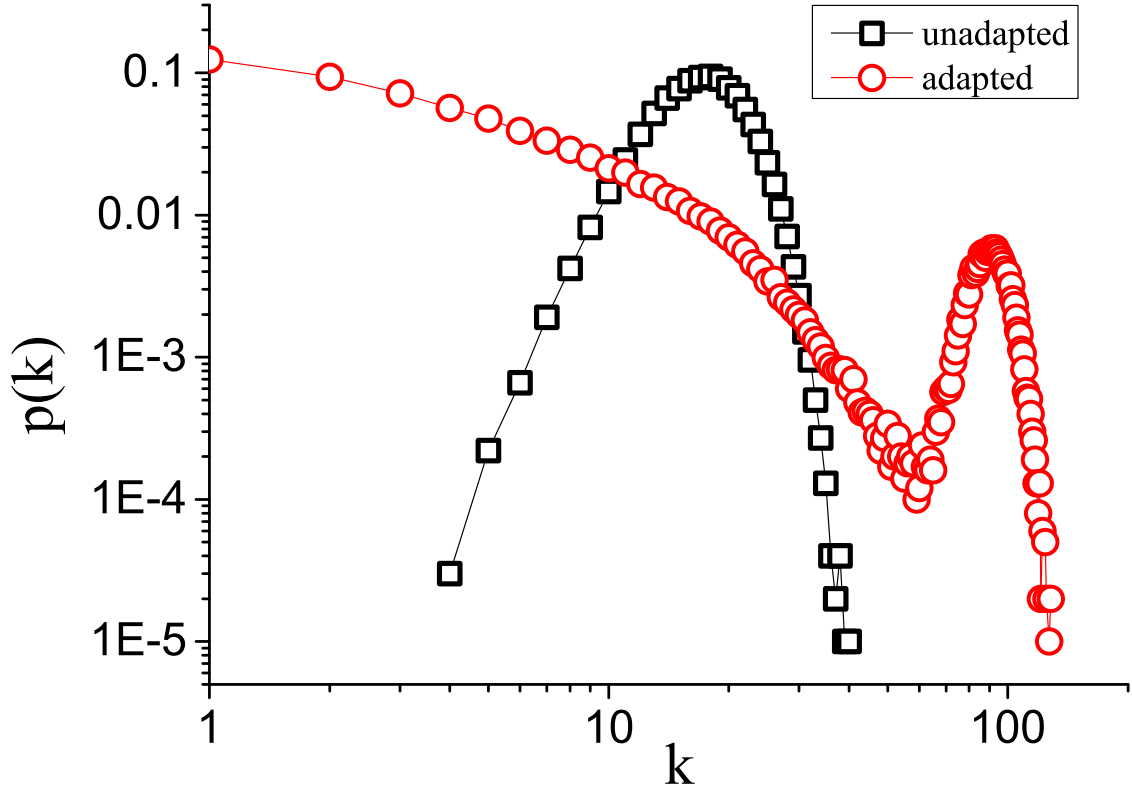,width=1\linewidth}
\caption{(Color online) {\bf Self-organized heterogeneity.} Comparison of the degree distributions of the initial unadapted network used in the first set of simulations and the `adapted' self-organized network. The self-organized network is significantly more heterogeneous. However, at the same time it is more resistant to disease invasion. Parameters: $\beta = 0.032$, $\psi_{\rm a} = 0.65$, $\psi_{\rm b} = 0.05$, $\omega=0.2$, $ \mu=0.002 $, $ N=10^{5} $, $ K=10^{6} $.}\label{Fig5}
\end{center}
\end{figure}

\emph{\textbf{Higher thresholds in heterogeneous networks}}.
Let us emphasize that the adapted network has a more heterogeneous degree distribution (Fig.~\ref{Fig5}) but also a higher epidemic threshold. Thus comparing the adapted and the maximally random state, the more heterogeneous network is actually more robust to the invasion of the pathogen. This appears counter-intuitive in the light of many recent work in network epidemiology, which showed that more heterogeneous contact networks are generally more susceptible to disease invasion. However, it was already shown in \cite{smilkov2014beyond} that heterogeneous networks become less susceptible if there is a correlation such that the highly connected nodes have lower susceptibility to the disease. The present results show that a simple but plausible local adaptation rule can drive this process so far that the network with heterogeneous degree distribution is more robust against disease invasion than an Erd\H{o}s-Renyi random graph of the same mean degree. 

Link heterogeneity is clearly a double-edged sword. On the one hand it is intuitive that some amount of heterogeneity in the degree is advantageous if it means that more links lead to nodes with low susceptibility. On the other hand, epidemic theory suggests that in the limit of very heterogeneous networks, the robustness of the network should decline because of the high excess degree\cite{pastor2001epidemic, moreno2002epidemic}. This suggests that there should be an optimum level of heterogeneity, where the epidemic invasion threshold is highest.

We can understand the interplay between the two effects by a link-centric percolation argument. We consider the limit of low disease prevalence and focus on the active links, i.e.~links connecting an infected node to a susceptible node. Given a single focal active link we can estimate the expected number of secondary active that is created by transmission along the focal link 
by 
\begin{equation}
Z_0=\frac{p_{\rm a}k_{\rm a}^2\psi_{\rm a} \beta}{\mu \bar{k}}+\frac{p_{\rm b}k_{\rm b}^2\psi_{\rm b} \beta}{\mu \bar{k}}
\end{equation}
where $\bar{k}=p_{\rm a}k_{\rm a}+p_{\rm b}k_{\rm b}$ is the constant mean degree of the system. 
Defining $q=k_{\rm b}/k_{\rm a}$ and substituting $k_{\rm a}=\bar{k}/(p_{\rm a} +p_{\rm b}q)$ and 
$k_{\rm b}=\bar{k}/(p_{\rm a}q^{-1} +p_{\rm b})$ we can express the link reproductive number $Z_0$ as a function 
of the degree ratio $q$. Although the resulting expression for $Z_0$ is relatively complex, we can find it's minimum, 
i.e.~the most robust point, by differentiating, which yields
\begin{equation}
\frac{k_{\rm b}}{k_{\rm a}}=\frac{\psi_{\rm a}}{\psi_{\rm b}}
\end{equation}
This coincides with the upper bound for the degree ratio, or, in other words, the point where the nodes of types A and B 
are infected at an identical rate. Therefore increasing the degree heterogeneity by increasing the degree ratio is advantageous 
to the point where the more resistant agents become infected more often than their less resistant counterparts. Thus it appears that the decisive characteristic that determines its robustness to disease invasion is not the heterogeneity of the the network structure, but the heterogeneity of the disease risk to which the agents are exposed.

The independence of the result above from other parameters suggests that it is true in a wider class of systems, but this intuition will need to be validated in further investigations.
For the adaptive system this result means that the network always operates in the regime where a higher degree ratio and 
therefore more heterogeneity has a stabilizing effect. We could in principle construct a system that self-organizes to the 
optimal point, by replacing the per-link rewiring rate by a rewiring rate per infected node. 
However this variant of the model is beyond the scope of the present paper.  

\section*{Discussion} \label{sec:Discussion}

In this paper we studied an adaptive heterogeneous SIS model numerically and analytically.
The analysis revealed that heterogeneity in the intrinsic parameters of the nodes induces heterogeneity 
in the connectivity: Over time more resistant nodes gain more links until a steady state 
is reached, in which nodes with higher resistance are still less likely to contract the disease, but more 
highly connected than average nodes. 

A well known result in network science is that more heterogeneous networks are less resistant to the invasion of
diseases. However, in the self-organized networks studied here the opposite effect is observed. 
In comparison to random networks the self-organized networks are both more resistant to the disease and more 
heterogeneous in connectivity. 

The evolved networks gain their resistance to the disease from correlations between intrinsic parameters  
and the node connectivity. The increased resistance thus arises directly from the heterogeneity. A comparable effect 
is not possible in networks of identical nodes. While the specific bifurcation structure of the present model may depend on modelling assumptions, the basic interplay between the connectivity and the invasion threshold arises from the fundamental physics of the spreading process and is thus likely to be a generic feature that is observed across many models.   
It thus appears plausible that also in real world epidemics some anti-correlation between the true susceptibility and the effective degree of agents will be induced. By concentrating more of the remaining links of the more resilient individually the network will generally become both more heterogeneous and more resistant to the disease.

It is possible that in real networks the self-organized heterogeneity is relatively minor compared to network heterogeneity that is unrelated to the epidemic in question. In that case the phenomenon described here would still occur but be of lesser importance. The extent to which self-organized heterogeneity plays a role in real world diseases will certainly depend on the disease in question, and can probably only be assessed through further empirical work. 

For epidemiology the results reported here imply, that network heterogeneity should not be considered in isolation. 
If there is an underlying heterogeneity in the susceptibility to the disease then a heterogeneous network may be more 
resistant to disease invasion than its homogeneous counterpart. Moreover, a vaccination campaign that targets the 
most highly connected nodes may end up vaccinating the wrong people as these nodes may also have the strongest natural resistance against the disease.

Perhaps most importantly, our results suggest that the invasibility of the network is not governed by the heterogenity of the network alone, but by the heterogeneity of effective disease risk, which takes both node degree and susceptibility into account. 

\section*{Methods}
\emph{\textbf{Moment-closure approximation.}}
The time evolution of the proportion of nodes that are infected and of type $u$ can then be written as 
\begin{equation}\label{eq_node_density}
\frac{\rm d}{\rm dt}[I_u]=-\mu [I_u]+\beta \psi_{u} \sum_v [S_uI_v]
\end{equation}     
The two terms of this equation capture the recovery of infected nodes and the infection of susceptible nodes, respectively. 
In the second term we used the symbol $[S_uI_v]$ to denote the number of links between a node of state S and type $u$ and a node of state I and type $v$, normalized with respect to the total number of nodes. This quantity therefore has the dimension of links per node. For determining these \emph{link densities}, we can write additional evolution equations, which are in turn depend on the density of triplet chains of nodes of given type and state $[X_uY_vZ_w]$, this yields 
\begin{equation}\label{eq_link_density_1}
\begin{split}
\frac{\rm d}{\rm dt}[S_uS_v] = & \kappa_{1} \mu([S_uI_v]+[I_uS_v])- \kappa_{1} \beta\sum_{w}(\psi_{v}[S_uS_vI_w]+\psi_{u}[S_vS_uI_w]) \\
&+ \frac{\kappa_{1} \omega}{\sum_{w}S_w}(\sum_w [S_v][S_uI_w]+[S_u][S_vI_w]),
\end{split}
\end{equation}
\begin{equation}\label{eq_link_density_2}
\begin{split}
\frac{\rm d}{\rm dt}[I_uI_v] = & -2\mu[I_uI_v]+\kappa_{1}\beta(\psi_{v}[I_uS_v]+\psi_{u}[S_uI_v]) \\
& + \kappa_{1} \beta (\psi_{v}\sum_{w}[I_uS_vI_w]+\psi_{u}\sum_{w}[I_vS_uI_w]),
\end{split}
\end{equation}
\begin{equation}\label{eq_link_density_3}
\begin{split}
\frac{\rm d}{\rm dt}[S_uI_v]= & \kappa_{2}\mu[I_uI_v]-(\mu+\beta\psi_{u})[S_uI_v] \\
&+\beta\psi_v\sum_{w}[S_uS_vI_w]-\beta\psi_u\sum_{w}[I_vS_uI_w]-\omega[S_uI_v],
\end{split}
\end{equation}
where 
\begin{equation}
\kappa_{1}=\begin{cases} 1/2 & u=v \\ 1 & u \neq v \end{cases}\,\,\mbox{and}\,\,\kappa_{2}=\begin{cases} 2 & u=v \\ 1 & u \neq v \end{cases}.
\end{equation}

In the equation for $[S_uS_v]$, the S-S-links, the first term $\kappa_{1}\mu([S_uI_v]+[I_uS_v])$ accounts for the creation of $S_u$-$S_v$-links by recovery of an infected node in an S-I-link. The factor $\kappa_1$ needs to be included to avoid double counting in the case of identical indices. The second term accounts for the destruction of $S_u$-$S_v$-links due to infection of one of the two S-nodes by a node that is external to the link. The number of such infected nodes outside the S-S-link is given by the number of triplets $[S_uS_vI_w]$ and $[S_vS_uI_w]$. For a single $S_a$-$S_b$-link the infection rate due to infected connected to the $S_b$-node is $\beta \psi_{\rm b} ([S_{\rm a}S_{\rm b}I_{\rm a}]+[S_{\rm a}S_{\rm b}I_{\rm b}])/[S_{\rm a}S_{\rm b}]$, i.e.~the expected number of triplet chains that run through one specific the S-S-link and include an infected node on the $S_a$ side, multiplied by the effective infection rate $\beta\psi_b$. To obtain the total rate we multiply by the number of those links, $[S_aS_b]$, which cancels the denominator, take the infected on the other side of the S-S-link into account, and multiply $\kappa_1$ to avoid double-counting. The final term in the equation accounts for creation of S-S-links by rewiring. Links of the type $S_a$-$I_b$ are rewired at rate $\omega$. When rewiring the susceptible node cuts the link to the infected and connects to a randomly chosen susceptible. In such a rewiring event, a new $S_a$-$S_a$-link is created if the newly-chosen partner is of type A. This is the case with probability $[S_a]/([S_a]+[S_b])$. Such that the total rate of $S_a$-$S_a$-link creation from $S_a$-$I_b$-link rewiring is $\omega[S_aI_b][S_a]/([S_a]+[S_b])$. Taking all possible combinations of indices and double-counting into account leads to the term in the equation. 

In the equation for $I_u$-$I_v$-links the first term accounts for the loss of these links due to recovery of one of the linked nodes. The second term, $\kappa_1\beta(\psi_{v}[I_uS_v]+\psi_{u}[S_uI_v])$, accounts for the creation of these links due to transmission of infection inside an S-I-link. Similarly, the final term accounts for the creation of I-I-link by infection of the S-node in an S-I-link from an internal source. In this term triplet chains appear in analogy to the term for the loss of S-S-links due to infection from sources external to the link, discussed above. In the equation for the S-I-links, the first term accounts for the creation of these links due to recovery of one infected node in an I-I-link. The second term accounts for the loss of S-I-links, both due to recovery of the I-node and due to internal transmission of the infection, resulting in an I-I-link. The fourth term captures the creation of S-I-links due to infection of an S-node in an S-S-link, whereas the fourth term captures the loss of S-I-links due to infection of the S-node from a source external to the link. Finally, the last term accounts for the loss of S-I-links due to rewiring.        

To cut the progression to ever larger network motifs one approximates the density of triplet chains by a moment closure approximation, here the pair approximation 
\begin{equation}\label{eq_pair_approximation}
[X_uY_vZ_w]=\delta\frac{[X_uY_v][Y_vZ_w]}{[Y_v]}
\end{equation}
where $\delta$ is a factor arising from symmetries, such that $\delta=4$ if $X_u=Y_v=Z_w$, $\delta=2$ if either $X_u=Y_v\neq Z_w$ or $X_u\neq Y_v=Z_w$, and $\delta=1$ if $X_u\neq Y_v\neq Z_w$.

Inherent in the moment-closure approximation is the assumption that long-ranged correlations vanish, such that the densities of motifs beyond the cut-off conform to statistical expectations. This assumption is the main source of inaccuracies in this type of approximation \cite{demirel2014moment}. The approximation can still be used to identify phase transitions in the adaptive SIS model as the correlations associated with these are still captured. However, the approximation performs poorly in fragmentation-type transitions, found for instance in the adaptive voter model\cite{bohme2011analytical}.

The additional equations from the moment closure approximation that are used in the continuation are 
\begin{equation}
\begin{split}
\frac{d[S_{\rm b}S_{\rm b}]}{dt}= & \mu [S_{\rm b}I_{\rm b}] -2\beta\psi_{\rm b} (\frac{[S_{\rm b}S_{\rm b}][S_{\rm b}I_{\rm a}]}{[S_{\rm b}]}+\frac{[S_{\rm b}S_{\rm b}][S_{\rm b}I_{\rm b}]}{[S_{\rm b}]}) 
+\frac{\omega [S_{\rm b}]}{[S_{\rm a}]+[S_{\rm b}]}([S_{\rm b}I_{\rm a}]+[S_{\rm b}I_{\rm b}]),
\end{split}
\end{equation}
\begin{equation}
\begin{split}
\frac{d[S_{\rm a}S_{\rm b}]}{dt}=&\mu ([S_{\rm b}I_{\rm a}]+[S_{\rm a}I_{\rm b}]) 
-\beta\psi_{\rm a} (\frac{[S_{\rm b}S_{\rm a}][S_{\rm a}I_{\rm a}]}{[S_{\rm a}]} +\frac{[S_{\rm b}S_{\rm a}][S_{\rm a}I_{\rm b}]}{[S_{\rm a}]}) 
-\beta\psi_{\rm b} (\frac{[S_{\rm a}S_{\rm b}][S_{\rm b}I_{\rm a}]}{[S_{\rm b}]} \\
&+\frac{[S_{\rm a}S_{\rm b}][S_{\rm b}I_{\rm b}]}{[S_{\rm b}]}) 
+\frac{\omega [S_{\rm b}]}{[S_{\rm a}]+[S_{\rm b}]}([S_{\rm a}I_{\rm a}]+[S_{\rm a}I_{\rm b}]) 
+\frac{\omega [S_{\rm a}]}{[S_{\rm a}]+[S_{\rm b}]}([S_{\rm b}I_{\rm a}]+[S_{\rm b}I_{\rm b}]),
\end{split}
\end{equation}

\begin{equation}
\begin{split}
\frac{d[S_{\rm a}I_{\rm a}]}{dt}=&2\mu [I_{\rm a}I_{\rm a}]-(\mu +\beta \psi_{\rm a} + \omega)[S_{\rm a}I_{\rm a}]
+2\beta\psi_{\rm a}(\frac{[S_{\rm a}S_{\rm a}][S_{\rm a}I_{\rm a}]}{[S_{\rm a}]}+\frac{[S_{\rm a}S_{\rm a}][S_{\rm a}I_{\rm b}]}{[S_{\rm a}]})\\
&-\beta\psi_{\rm a}(\frac{[S_{\rm a}I_{\rm a}][S_{\rm a}I_{\rm a}]}{[S_{\rm a}]}
+\frac{[S_{\rm a}I_{\rm a}][S_{\rm a}I_{\rm b}]}{[S_{\rm a}]}),
\end{split}
\end{equation}
\begin{equation}
\begin{split}
\frac{d[S_{\rm b}I_{\rm b}]}{dt}=&2\mu [I_{\rm b}I_{\rm b}]-(\mu +\beta \psi_{\rm b} + \omega )[S_{\rm b}I_{\rm b}] 
+2\beta\psi_{\rm b}(\frac{[S_{\rm b}S_{\rm b}][S_{\rm b}I_{\rm a}]}{[S_{\rm b}]}+
\frac{[S_{\rm b}S_{\rm b}][S_{\rm b}I_{\rm b}]}{[S_{\rm b}]})\\
&-\beta\psi_{\rm b}(\frac{[S_{\rm b}I_{\rm b}][S_{\rm b}I_{\rm a}]}{[S_{\rm b}]}
+\frac{[S_{\rm b}I_{\rm b}][S_{\rm b}I_{\rm b}]}{[S_{\rm b}]}),
\end{split}
\end{equation}
\begin{equation}
\begin{split}
\frac{d[S_{\rm a}I_{\rm b}]}{dt}=&\mu [I_{\rm a}I_{\rm b}]-(\mu +\beta \psi_{\rm a} + \omega)[S_{\rm a}I_{\rm b}] 
+\beta\psi_{\rm b}(\frac{[S_{\rm a}S_{\rm b}][S_{\rm b}I_{\rm a}]}{[S_{\rm b}]}+
\frac{[S_{\rm a}S_{\rm b}][S_{\rm b}I_{\rm b}]}{[S_{\rm b}]})\\
&-\beta\psi_{\rm a}(\frac{[S_{\rm a}I_{\rm b}][S_{\rm a}I_{\rm a}]}{[S_{\rm a}]}
+\frac{[S_{\rm a}I_{\rm b}][S_{\rm a}I_{\rm b}]}{[S_{\rm a}]}),
\end{split}
\end{equation}
\begin{equation}
\begin{split}
\frac{d[S_{\rm b}I_{\rm a}]}{dt}=&\mu [I_{\rm a}I_{\rm b}]-(\mu +\beta \psi_{\rm b} + \omega)[S_{\rm b}I_{\rm a}] 
+\beta\psi_{\rm a}(\frac{[S_{\rm a}S_{\rm b}][S_{\rm a}I_{\rm a}]}{[S_{\rm a}]}+
\frac{[S_{\rm a}S_{\rm b}][S_{\rm a}I_{\rm b}]}{[S_{\rm a}]})\\
&-\beta\psi_{\rm b}(\frac{[S_{\rm b}I_{\rm a}][S_{\rm b}I_{\rm a}]}{[S_{\rm b}]}
+\frac{[S_{\rm b}I_{\rm a}][S_{\rm b}I_{\rm b}]}{[S_{\rm b}]}),
\end{split}
\end{equation}

\begin{equation}
\begin{split}
\frac{d[I_{\rm a}I_{\rm a}]}{dt}= & -2\mu [I_{\rm a}I_{\rm a}]+\beta \psi_{\rm a}[S_{\rm a}I_{\rm a}] 
+\beta\psi_{\rm a}(\frac{[S_{\rm a}I_{\rm a}][S_{\rm a}I_{\rm a}]}{[S_{\rm a}]}+
\frac{[S_{\rm a}I_{\rm a}][S_{\rm a}I_{\rm b}]}{[S_{\rm a}]}),
\end{split}
\end{equation}
\begin{equation}
\begin{split}
\frac{d[I_{\rm b}I_{\rm b}]}{dt}= & -2\mu [I_{\rm b}I_{\rm b}]+\beta \psi_{\rm b}[S_{\rm b}I_{\rm b}] 
+\beta\psi_{\rm b}(\frac{[S_{\rm b}I_{\rm b}][S_{\rm b}I_{\rm a}]}{[S_{\rm b}]}+
\frac{[S_{\rm b}I_{\rm b}][S_{\rm b}I_{\rm b}]}{[S_{\rm b}]}).
\end{split}
\end{equation}


\section*{Acknowledgement}

This work was supported by the EPSRC under grant EP/K031686/1, National Natural Science Foundation of China (Grant Nos. 11105025, 91324002) and the Program of Outstanding Ph.D. Candidate in Academic Research by UESTC (Grant No. YBXSZC20131036).

\section*{Author contributions}
H. Y., M. T. and T. G. devised the research project.
H. Y. performed numerical simulations.
H. Y. and T. G. analysed the results.
H. Y.,  M. T. and T. G. wrote the paper.

\section*{Additional information}



{\bf Competing financial interests}:
The authors declare no competing financial interests.

\section*{Figure legends}

\textbf{Figure 1:} {\bf Bifurcation diagram of the adaptive heterogeneous SIS model.} Show is the stationary level of disease prevalence $I^*$ as a function of infectivity $\beta$. When the infectivitiy is decreased the endemic state vanishes in a saddle node bifurcation (`Persistence threshold'). The disease free state can be invaded by the epidemic, if the infectivity surpasses a point where a transcritical bifurcation occurs (`Invasion threshold'). Agent-based simulation (circles) and equation-based continuation (lines) provide consistent results on the persistence threshold, but predict different invasion thresholds. This discrepancy appears due to a projection effect, because the state where the disease is extinct is not uniquely defined (see text). In addition to stable solution branches (solid) the continuation also reveals an unstable solution branch (dotted). Parameters:  $ \psi_{\rm a} = 0.65 $, $ \psi_{\rm b} = 0.05 $, $ p_{\rm a}= 0.75 $, $ \mu=0.002 $, $ N=10^{5} $, $ K=10^{6} $.

\textbf{Figure 2:}  {\bf Comparison of thresholds.} The plot shows a very good agreement agreement between equation-based continuation (lines) and agent-based simulations (symbols) for the persistence thresholds $\beta_{\rm per}$ (box, dashed). However, a notable difference exists for the invasion thresholds $\beta_{\rm inv}$ (circle, dotted). Parameters: $ \psi_{\rm b} = 0.05 $, $ \mu=0.002 $, $ N=10^{5} $, $ K=10^{6} $.

\textbf{Figure 3:} {\bf Network heterogeneity in the adapted state, indicated by the degree ratio $k_{\rm b}/k_{\rm a}$.} The dependence of the degree ratio in agent-based simulations (black circles) closely follows the prediction from integration of the ODE model (solid red line), a very good approximation is also provided by the relationship $k_{\rm b}/k_{\rm a} = \sqrt{\psi_{\rm a}/\psi_{\rm b}}$ (blue dashed line), whereas the naive expectation $k_{\rm b}/k_{\rm a} = \psi_{\rm a}/\psi_{\rm b}$ (pink dotted line) overestimates the network heterogeneity significantly. Contrary to expectations the networks following the naive solution (the most heterogeneous case), would be maximally stable against disease invasion. Parameters: $ \psi_{\rm b} = 0.05 $, $ \mu=0.002 $, $ N=10^{5} $, $ K=10^{6} $. Inset: the magnification of the superimposed part of the main figure.

\textbf{Figure 4:} {\bf Comparison of thresholds in self-organized networks.} The plot shows a very good agreement between equation-based continuation (lines) and agent-based simulations started in an artificially created adapted state (symbols) for both the invasion thresholds $\beta_{\rm inv}$ (circle, dotted) and the persistence thresholds $\beta_{\rm per}$ (box, dashed). See Fig.~2 for comparison. Parameters: $ \psi_{\rm b} = 0.05 $, $ \mu=0.002 $, $ N=10^{5} $, $ K=10^{6} $.

\textbf{Figure 5:} {\bf Self-organized heterogeneity.} Comparison of the degree distributions of the initial unadapted network used in the first set of simulations and the `adapted' self-organized network. The self-organized network is significantly more heterogeneous. However, at the same time it is more resistant to disease invasion. Parameters: $\beta = 0.032$, $\psi_{\rm a} = 0.65$, $\psi_{\rm b} = 0.05$, $ \mu=0.002 $, $ N=10^{5} $, $ K=10^{6} $.

\end{document}